# CEGANN: Crystal Edge Graph Attention Neural Network for multiscale classification of materials environment

Suvo Banik[1,2], Debdas Dhabal[3], Henry Chan[1], Sukriti Manna[1,2], Mathew Cherukara[4], Valeria Molinero[3], Subramanian KRS Sankaranarayanan[1,2]

[1] Center for Nanoscale Materials, Argonne National Laboratory, Lemont, Illinois 60439, United States.
[2] Department of Mechanical and Industrial Engineering, University of Illinois, Chicago, Illinois 60607, The United States.
[3] Department of Chemistry University of Utah Salt Lake City, UT 84112, United States.
[4] Advanced Photon Source, Argonne National Laboratory, Lemont, Illinois 60439, United States.

* Corresponding author: skrssank@uic.edu

**Abstract**:

Machine learning (ML) models and applications in materials design and discovery typically involve the use of feature representations or "descriptors" followed by a learning algorithm that maps them to a user-desired property of interest. Most popular mathematical formulation-based descriptors are not unique across atomic environments or suffer from transferability issues across different application domains and/or material classes. In this work, we introduce the Crystal Edge Graph Attention Neural Network (CEGANN) workflow that uses graph attention-based architecture to learn unique feature representations and perform classification of materials across multiple scales (from atomic to mesoscale) and diverse classes ranging from metals, oxides, non-metals and even hierarchical materials such as zeolites and semi ordered materials such as mesophases. We first demonstrate a case study where the classification is based on a global, structure-level representation such as space group and structural dimensionality (e.g., bulk, 2D, clusters etc.). Using representative materials such as polycrystals and zeolites, we next demonstrate the transferability of our network in successfully performing local atom-level classification tasks, such as grain boundary identification and other heterointerfaces. We also demonstrate classification in (thermal) noisy dynamical environments using a representative example of crystal nucleation and growth of a zeolite polymorph from an amorphous synthesis mixture. Finally, we characterize the formation of a binary mesophase and its phase transitions and the growth of ice, demonstrating the performance of CEGANN in systems with thermal noise and compositional diversity. Overall, our approach is agnostic to the material type and allows for multiscale classification of features ranging from atomic-scale crystal structures to heterointerfaces to microscale grain boundaries.

**Keywords**: Multiscale material classification, Machine Learning, Graph Attention Network, Crystal Graphs, Nucleation and Growth

**Introduction**:

Characterization of materials with unique properties[1–5] is at the core of data-driven material design and discovery[6,7]. A relatively small fraction of materials has been characterized either experimentally or with computational methods, compared to their anticipated potential diversity across a vast chemical space. Given the surge in the development of materials databases[8–10] in recent years, there is an urgent need for automated tools to analyze large amounts of structural data. In this regard, distinguishing the unique characteristics across different classes of materials with varying dimensionality can provide key insights into learnable aspects which are crucial for state-of-the-art machine-learning (ML) tools to be successfully implemented in the design and discovery of new materials with unique properties. To achieve such distinction, ML models typically involve the use of fingerprints or descriptors[11–16] that allow a learning algorithm to map the fingerprint to a user-desired property of interest. A descriptor that maps the crystal features in a vector space should always be (1) invariant to basis symmetries such as rotation, reflection, translation, and permutation of atoms[12], (2) unique to the system applied, but sensitive towards variation in properties, and (3) simple and robust. Additionally, these features play a crucial role in a plethora of applications such as quantitative structure-property relationship (QSPR)[11,17–20], development of interatomic potentials [13,21–24], and prediction of atomistic configurations based on targeted properties[25–29], surface phenomena[30].

A feature representation is constructed primarily in two ways (i) using a predefined mathematical formulation, or (ii) learning the representation by combining fundamental low-level features and correlating them to the relevant task being performed using ML methods. A plethora of mathematical formulation-based descriptors[12–15,18,31–35] such as radial distribution functions (RDF), angular distribution function (ADF), common neighbor analysis (CNA)[31], adaptive CNA[31], Centro symmetry parameter (CSP)[31], Voronoi analysis[31], Steinhardt order parameter (SP)[32], bond angle analysis (BAA)[33], and neighbor distance analysis (NDA)[31] are widely used for featurization. A majority of these are very simple and of a very local nature, *i.e.,* mostly capable of differentiating ordered and disordered structures. Improving upon these, a set of features can be developed using pairwise feature matrices and their transformations[12,18–20,34,35]. These features may be as simple as pairwise distances e.g., Weyl matrices[35], Z-matrices[36], or pairwise electrostatic interactions between atoms (Coulomb matrix[18], sine matrix[19]). A more comprehensive representation of these matrices is permutation histograms[15] e.g., MBTR (Many Body Tensor Representation)[12,20], BOB (Bag of Bonds). The advantages of these methods are that the pairwise features are translationally invariant, and these matrices present a unique representation of the system. However, a major setback of these matrix representations is that they are not invariant to changes in atom ordering. A very popular approach involves the use of smooth overlap of atomic positions (SOAP)[14] descriptors constructed by expanding the atomic neighbor density "$\rho$" on a spherical harmonics basis and then further expanding it on a radial basis to obtain the rotational invariant power spectrum[34]. SOAP provides a very robust representation of the local environment in a smooth and continuous manner which makes it very suitable for mapping potential energy surfaces. Nevertheless, most of the existing approaches for structure characterization that involve global comparison between two structures are either based on a simple aggregation-based method (average kernel) that causes loss in resolution or are computationally expensive such as, for example, the best match kernel method[14]. Another setback is that the number of descriptors increases quadratically[6] with the increase in the number of chemical species, precluding their applicability to multicomponent systems.

In this context, graph neural networks (GNNs)[37–39] have been widely used in node-level as well as graph-level classification tasks with remarkable success. Furthermore, recent developments in the area of graph attention networks[40] make the task more accurate by learning the interaction between node-level features. These networks tend to learn flexible representations by combining very fundamental low-level features (interatomic distances, bond angles, etc.), and yet produce a graph-based input that very accurately maps to the target-specific tasks. This caters to the limitations of static descriptors bound by predefined mathematical formulations.

In crystal systems, GNNs can be made to operate on atom-based graphs to create node-level embedding through convolutions on neighboring nodes and edges[41–48]. More layers of convolutions tend to capture higher-level information. A widely used framework for crystal systems is the Crystal Graph Convolutional Neural Network (CGCNN)[47,48]. Xie and Grossman have shown that CGCNN can directly learn material properties from the connectivity of atoms in a crystal, thus enabling an interpretable representation of crystalline materials[49]. Graph attention-based architecture[44] has also recently been implemented for the structure to property mapping in atomic systems. Traditional CGCNN architecture tends to map structure to property by using a diverse set of atom-level features (e.g., group number, period number, atomic number, electronic structure, *etc*.), and crystal graphs with simple edge feature such as pairwise interatomic distances.

Predefined mathematical formulation-based descriptors are useful when there isn't sufficient data to learn from. However, they largely suffer from transferability issues due to a lack of flexibility. On the other hand, current existing graph based CGCNN architectures do not incorporate orientational features[41,45] that are very relevant for classification tasks in a multitude of atomic environments. Moreover, these features tend to play a more significant role in classification tasks than features belonging to different atomic species. Although there have been recent applications[41,45] that include orientational features in their network architecture, they are more complex in nature and mostly focus on property prediction. To elucidate the issues involving transferability and applicability, we present two distinct classification scenarios (Fig. 1). To start with, we classify the liquid and glassy-amorphous phases of a representative material such as silicon (Si). Both liquid and amorphous phases are disordered with no symmetry whatsoever and only differ in density and coordination number. Fig. 1(a) displays the variability in the coordination environment of the liquids and amorphous phases used in this study. We use a dataset containing 2000 structures with 50% liquid and 50% amorphous (see supplementary information S.1 for the details of the data generation methods), and train a traditional CGCNN model using a train-to-validation split of 80:20 . From the t-SNE (t-distributed stochastic neighbor embedding) of feature representation of the validation dataset, with SOAP (cutoff 6 Å) (Fig. 1(b)) there is no distinct separation of the phases in features space indicating the inability of SOAP to distinguish the individual phases (see supplementary Fig. S1 (c) (d) for different cutoffs). On the other hand, simple bond order bases features ($Q_2$, $Q_4$, $Q_6$) + CN (Coordination Number) (cutoff 6 Å) Fig. 1(c) and a trained CGCNN (Fig. 1(d)) can clearly characterize the phases with decent separation in feature space. The second task involves the identification of particles belonging to local motifs (hexagonal or cubic) in a stacking disordered (ABABCB), ice. The correlated bond order-based CHILL+[50] is used as a benchmark for labeling the data (Fig. 1 (e)). Similar to the earlier case, we employ order parameters, SOAP, and CGCNN for this classification. The training data of CGCNN comprises a pure cubic, hexagonal, and stacking fault (ABCBCB) ice structure. The results in Fig. 1 (f) (g) (h) indicate that while SOAP is able to classify local motifs, CGCNN or the order parameters-based features fails to do so. Which is converse to the fact that SOAP couldn't characterize to structure belonging to liquid of amorphous class while its two counterparts could. This is a clear indication of the transferability issue in present characterization techniques across various problems at different scales. Despite traditional GNNs (such as CGCNN) showing exceptional promise in learning flexible feature representation at a graph level, (global) their performance in local environments is not as good as their global counterpart and remains mostly unexplored.

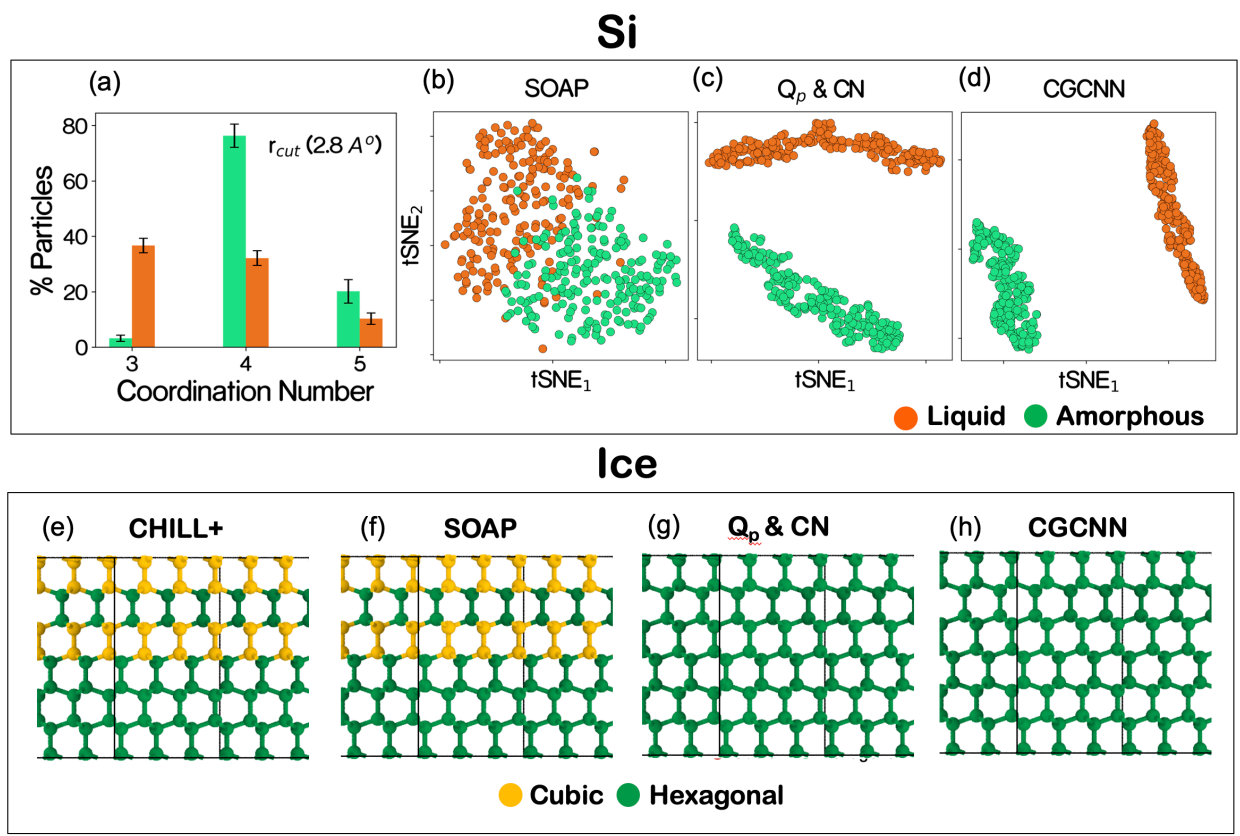

*Fig. 1.* *Classification of Silicon (Si) liquid and amorphous (glass) phases & atoms belonging to hexagonal and cubic motif in a ABABCB stacked ice. (a) Typical coordination number distribution of liquid and amorphous structures used for this study. (b) t-SNE plot of SOAP representation of the crystalline and amorphous phases in the validation dataset. (c) t-SNE plot for bond order parameters ($Q_2$, $Q_4$, $Q_6$) + CN (Coordination Number), feature representation. (d) t-SNE plot of embeddings of the validation dataset obtained training a classic CGCNN model. (e) Shows the classification of hexagonal and cubic stacked atoms in a ABABCB stacked ice with CHILL+[37] algorithm. (f) by using SOAP feature vector (g) by using order parameters ($Q_2$, $Q_4$, $Q_6$) + CN (Coordination Number) and (h) from the trained CGCNN model.*

Clearly, there is a need for a method that is not only transferrable but adaptable to variabilities in the material environment while providing accurate characterization at different scales. To the best of our knowledge, most efforts on crystal graph neural networks have been restricted to mapping structures to properties and a few property-based prediction tasks. There is still an immense untapped potential of GNNs in classification at both structure (global) and atomic (local) levels. In this work, we introduce a graph attention-based[51] workflow that operates on edge graphs, convoluting edges and bond angle features and passing messages in between (Fig. 2), to learn feature representation of material environments. A great advantage of attentional-based architectures is that they can learn the importance of feature vectors (i.e., bonds, angles) in a given neighborhood of each atom and put emphasis on ones unique to the task being performed. This helps in increasing the performance by ignoring redundant and unnecessary information. We demonstrate the efficacy of our workflow in classification tasks at both the atom-level (local) and structure-level (global) using a wide range of representative examples from materials applications. For global level classification, we perform two tasks. The first is classifying a diverse range of materials based on their space groups, and the second is classifying them based on their dimensionality (bulk, 2D, cluster etc.). We base the local atom level classification on structural motifs (FCC, BCC, HCP, diamond cubic), and demonstrate its use on a classic problem of grain boundary identification and grain size distribution. To validate the efficacy of our workflow in environments with thermal variations or noise, we deploy our classification workflow to facilitate the study of nucleation and growth of a zeolite, a complex porous crystal, in molecular dynamics simulations of synthesis. Often, practical materials application involves characterization of phases with structural and compositional variances along with thermal noise. We address these challenges through the identification of ice and liquid along simulations of water crystallization, and the classification of disordered, mesophase and crystalline order in binary mixtures in simulations involving transformations between these phases.

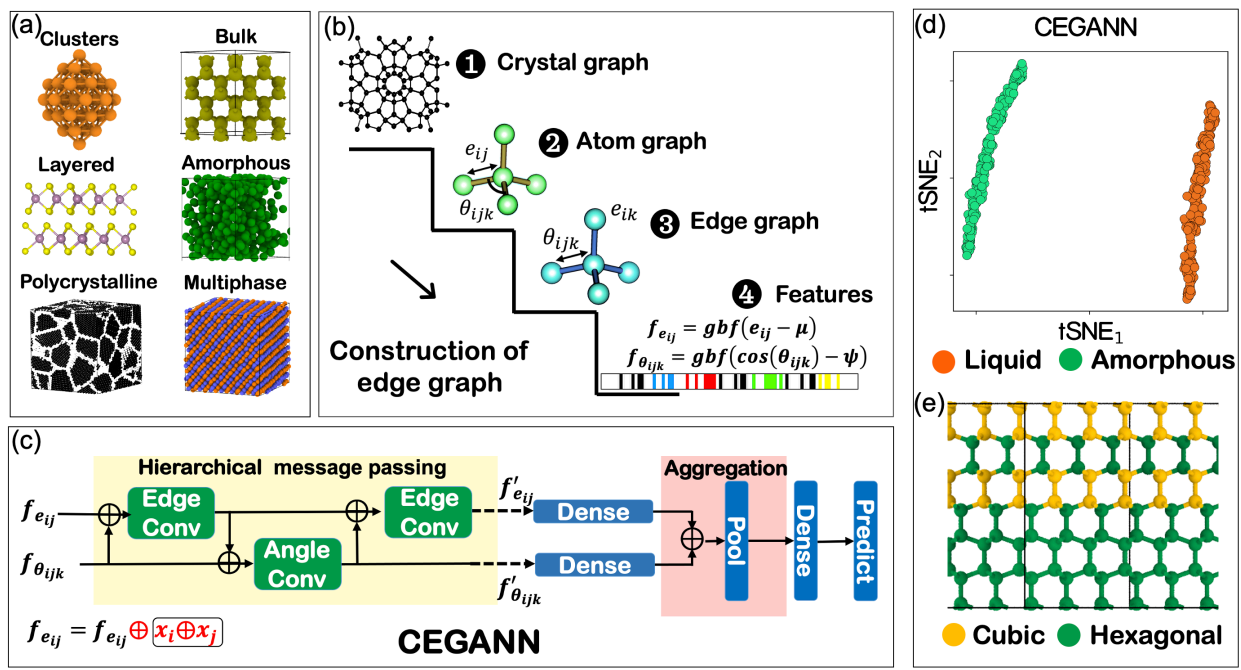

*Fig. 2.* Basic architecture of CEGANN workflow. Crystals are converted into atom graphs and edge graphs are obtained from atom graphs (a-b). (c) Shows alternate message passing and hierarchical interaction between edge and angle convolutions. Finally, the atom feature, convolved edge, and angle features are concatenated to produce the final representation. (d) t-SNE plot of the feature vector of liquid and amorphous structures as predicted by CEGANN workflow on an identical validation dataset as Fig. 1. (e) Shows identification of particles belonging to a hexagonal and cubic motif in an ABABCB stacked ice by a trained CEGANN workflow.

**Results**

**Edge graph representation**:
Edge graphs are higher-order representations of atomic graphs with edges as nodes and bond angles as connections between a pair of edges (Fig. 2(b)). We start from a crystal structure, creating its atom-graph (atom as nodes, bonds as edges) based on a fixed number of nearest neighbors. The edge graph is extracted from the atom-graph afterward (Fig. 2 (a) (b)). The edge features ($e_{ij}$) are obtained by expanding the pairwise distance on Gaussian basis functions while the bond angle features ($\theta_{ijk}$) are obtained by expanding the cosines of the bond angles on a gaussian basis as well.

**Hierarchical message passing:**

One main feature of the proposed architecture is the hierarchical interaction between edge and angle layers (Fig. 2(c)) (see methods section). The edge layer always gets updated first. This follows the hierarchy that the bond angles are constructed from a pair of edges and any change at the edge level should get updated first before passing the information onto the corresponding angle. Which gives **n-1** angle convolution operations for **n** edge convolutions where "**n**" is an integer.

**CEGANN Workflow for multiscale classification:**

The architecture of the CEGANN workflow used to perform multiscale classification of materials is shown in Fig. 2(c). The edge-graph feature representation of the structures is passed to the hierarchical message passing block for the convolution operations. The output of the convolved feature vectors from the edge and angle convolution layers are then passed to the aggregation block via dense layers (linear transformation), where feature representations of each of the structures are generated for the prediction task. For multicomponent system, additional chemical information can be included in the input edge feature vector $f_{e_{ij}}$(Fig. 2 (c)) as one hot encoding, depending on the characterization task being performed. CEGANN architecture also has an inherent ability of learning to distinguish atomic species from the interatomic distances of nearest neighbor atoms (see supplementary S.2, Fig. S2). The choice of the number of edges and angle convolution layers to be employed depends on the scale at which the classification tasks are being performed. For local-level tasks, it is preferable to have fewer convolutions while its global application requires more. In this work, we select an optimal number of convolutions that results in the best performance of our model for each of the tasks being performed (Table.1). Similar to the choice of the number of convolutions, the number of neighbors considered for the graph constructions also affect the model performance. A grid study can be performed to obtain an optimal set of hyperparameter for the specific task (see supplementary S.3, Fig. S3 (a-c)). In the end, the selection becomes entirely dependent on the choice of the problem, the computational cost associated and the accuracy of the prediction. The number of neighbors and the number of convolutions used for each of the tasks is reported in Table.1. It is to be noted that for all classification tasks performed in this work, we keep the input dimension of edge and angle feature vectors to be 80. We maintain uniformity of samples belonging to each class in both training and testing data while the splitting of any individual class is done randomly at a given ratio.

***Table. 1.*** *CEGANN Network hyperparameters used during different classification task*

| Classification task | Edge Convolution | Angle Convolution | Nearest neighbors |
|---|---|---|---|
| Amorphous and liquid | 2 | 1 | 12 |
| Stacking disordered ice | 1 | 0 | 16 |
| Space groups | 2 | 1 | 12 |
| Dimensionality classification | 2 | 1 | 12 |
| Grain size distribution (FCC) | 1 | 0 | 12 |
| Grain size distribution (BCC) | 1 | 0 | 14 |
| Grain size distribution (Diamond) | 1 | 0 | 16 |
| Grain size distribution (HCP) | 1 | 0 | 12 |
| Dynamical classification with noise | 1 | 0 | 12 |
| Mesophase chracterization | 1 | 0 | 12 |
| Interfacial growth of ice | 1 | 0 | 12 |

**Classification of liquid and amorphous silicon and stacking-disordered ice:**

We start by employing our CEGANN workflow for the classification tasks as discussed in Fig. 1. (a) Classification of liquid and amorphous phases (Silicon) Fig. 1 (a-d) (b) Characterization of local motifs (Hexagonal or Cubic) in stacking disordered ice (ABABCB) Fig. 1 (e-h). CEGANN is trained on the same training data as CGCNN (Fig. 1 (d), (h)). The validation data is also kept identical. From Fig. 2 (d) the t-SNE plot of the feature vectors of liquid and amorphous structures as predicted by CEGANN, it is evident that CEGANN has been able to distinguish amorphous and liquid phases of silicon conspicuously. Fig. 2 (e) also depicts the ability of CEGANN to precisely classify local cubic and hexagonal motifs in stacking fault structures where CGCNN performed miserably. CEGANN can overcome the challenge of transferability for both applications ranging from global to local levels while its counterparts, such as the traditional CGCNN, and descriptors such as SOAP fail to do so (Fig.1).

**Characterization of crystal structures based on their space groups:**

The space group of a crystalline system directly correlates to its structural motif, albeit at a global level. We demonstrate that the CEGANN framework can classify several different material classes based on their space groups. For this classification task, we use the same dataset as in Ref. 52. The space group of each crystal is calculated using a Pymatgen[52] package. The dataset contains a total of 10517 crystal structures with 7 crystal classes belonging to 8 different space groups. For the elemental system, the classes are body-centered tetragonal (bct, 139 and 141), rhombohedral (rh, 166), hexagonal (hex, 194), simple cubic (sc, 221), face-centered cubic (fcc, 225), diamond (dia, 227), and body-centered cubic (bcc, 229) respectively (see supplementary Fig. S4 (a)).

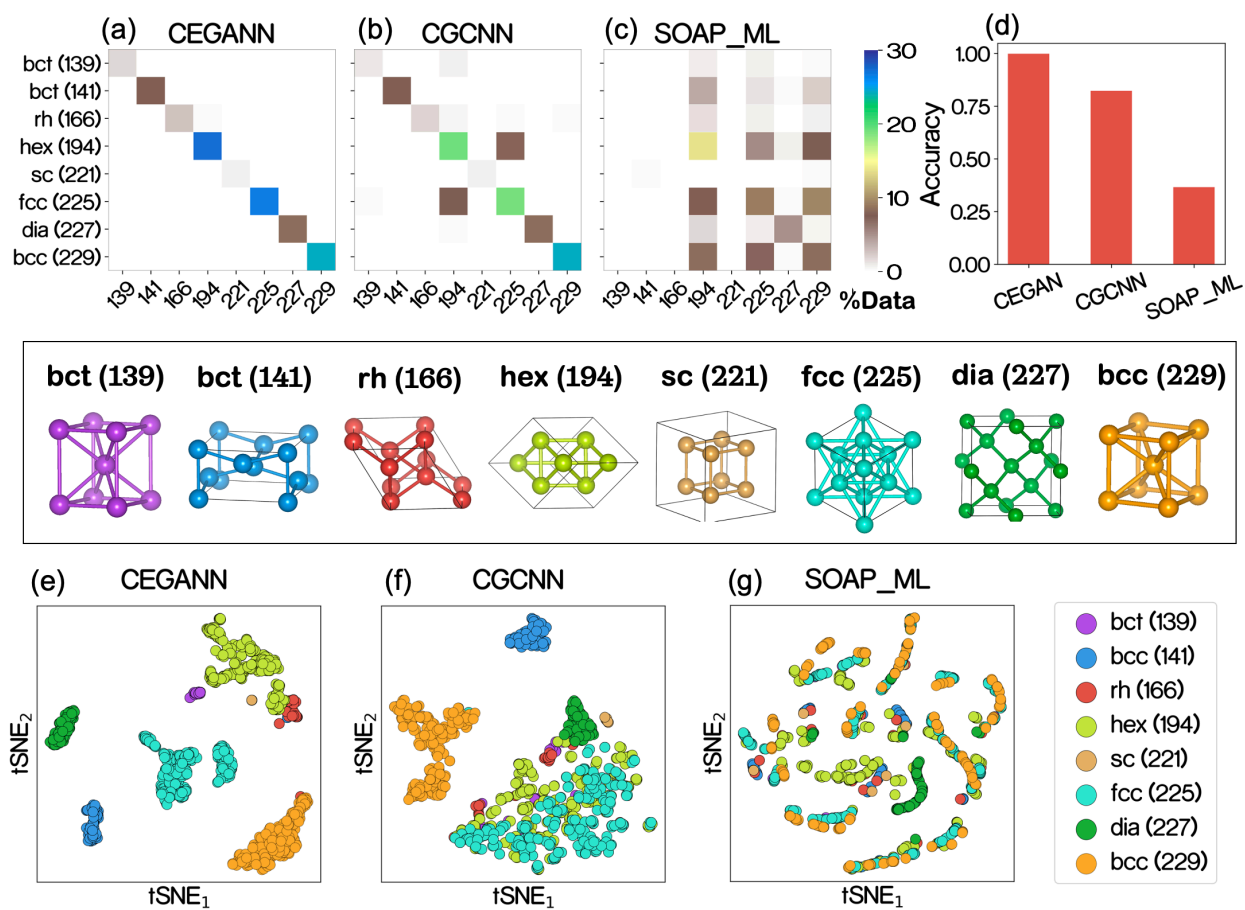

*Fig. 3.* *Global classification of crystals structures based on space groups. (a-c) shows the confusion matrix for CEGANN, CGCNN, SOAP ML, and workflow respectively. (d) Shows accuracy of prediction on the validation set for the 3 different architectures used. (e-g) show the t-SNE plot of the embeddings in feature space as learned by the CEGANN, CGCNN, and SOAP ML, respectively*

We start with the dataset having a train-to-validation ratio of 90:10 and train CEGANN, CGCNN, and a SOAP_ML workflow on this dataset. It is worth noting that our goal is to map SOAP feature vectors (cutoff 6 Å) directly to the space group. So, instead of passing the SOAP features through consecutive dense layers (linear transformation) with nonlinear activations[53], we have only one dense layer that directly maps it to the target space (SOAP_ML workflow) conforming to the specification used in CEGANN after the aggregation block (Fig. 2 (c)). The accuracy on the validation dataset is shown in Fig. 3 (d). The CEGANN workflow achieved an accuracy of ~100% on the validation set. The confusion matrix of CEGANN (Fig. 3(a)) also demonstrates a perfect identification (no off-diagonal entries) (also see supplementary Fig. S4 (b)-(i)) of each class belonging to different space groups. The CGCNN on the other hand achieves an accuracy of ~83% on the validation dataset with major confusion (Fig. 3 (b)) between the hex (194) and fcc (225) structure. This is very evident from the fact that fcc and hcp are close packed with a 74% atomic packing factor, and 12 nearest neighbors for both, which results in an identical graphical representation of the structures unless the orientational order of the particles are considered. The CGCNN not having these attributes in its graphical representation, significantly impacts its performance. The performance of SOAP_ML workflow is poor indicating that SOAP in its current mathematical state, however, does contain all the information but is not flexible enough to be directly mappable to the target space group. The degree of characterization can also be visualized in the t-SNE plot of the feature space representation on the validation dataset (Fig. 3 (e-g)). There is a clear distinction in the representation of each class for CEGANN, while CGCNN and SOAP feature vectors display a lack of resolution in the representation of each class in the feature space.

**Classification of polymorphs across various structural dimensionalities:**
Next, we demonstrate the ability of CEGANN to perform classification on material polymorphs across various dimensionalities, from clusters (0D) to sheets (2D) to bulk (3D). Carbon is known to have a diverse range of allotropes across these dimensionalities, making it an excellent candidate for validating the performance of our network for dimensionality classification. We start with a dataset of 511 bulk structures collected from the Samara Carbon Allotrope Database (SACADA)[54], Monolayer C polymorphs[55], and Graphite with varying interlayer distances, a collection of different Graphite allotrope and 2D polymorphs Carbon sampled using CASTING framework[56] and LCBOP potential[57] making a total of 612, 2D structures and 704 C nanoclusters[58,59] resulting in a total dataset of 1827 configurations (see supplementary Fig. S5). We divide our dataset into 80% training and 20% validation.

Fig. 4 (a) shows the confusion matrix for the dimensionality classification. CEGANN workflow can classify the structures with ~100% accuracy. Fig. 4 (b) shows the t-SNE plot of the embeddings of the validation set data. A clear distinction between phases can also be observed in the feature space which displays the capability of CEGANN to characterize polymorphs of different dimensions. It is worth mentioning that dimensionality is a defining material parameter, depending on which material can exhibit dramatically different properties[60]. Identification of materials based on their dimensionality is a crucial aspect of new material design and prediction[61]. While 3D crystalline objects are well documented among the experimentally known crystals, the same is not true for low dimensional structures such as 2D or 0D. For example, in a few cases where isolated 2D carbon layers tend to form porous bulk-like polymorphs which makes it difficult to categorize and distinguish them from typical layered structures.

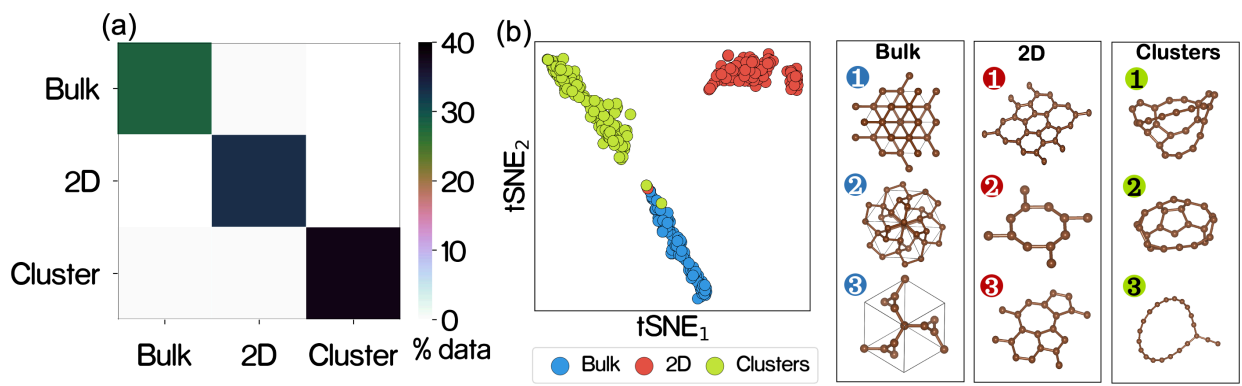

*Fig. 4.* *Classification of carbon polymorphs of different dimensionalities 0D (Clusters), 2D (Sheets), and 3D(Bulk). (a) Shows the confusion matrix of the prediction on the validation dataset by CEGANN. (b) The t-SNE plot of the feature representation of the validation dataset as predicted by CEGANN.*

**Grain boundary identification:**

Characterization of local motifs in full 3D samples of polycrystalline materials and accurately identifying grains and boundaries is a nontrivial task with a plethora of applications in material science. Although there are many methods used for grain characterization[31,33], there is no gold standard for identifying the grain size distribution in polycrystalline materials, as the predictions widely vary with the methodology used. We use CNA[31] (Common Neighbor Analysis) as a benchmark to generate labels for the training and validation data. CNA has been widely utilized for characterization of local motifs in ordered and disordered systems[62–65]. The original CNA method is based on the signatures of the local neighborhood of an atom and matching it to a reference one. The neighborhood of an atom is constructed based of a fixed cutoff ($r_{cut}$). The overall atomic signature of the atom consists of three features: (1) the number of neighbor atoms the central atom and its bonded neighbor have in common, $n_{cn}$, (2) the total number of bonds between these common neighbors, $n_b$, and (3) the number of bonds in the longest chain of bonds connecting the common neighbors, $n_{lcb}$. However, traditional CNA, and even its variations (such as adaptive CNA), not only show variability in results, but their performance also deteriorates under physical deformtions[65].

Here we consider for the prediction task 4 representative polycrystal classes (i) Face-Centered Cubic (FCC- Al), Body-Centered cubic (BCC-W), Diamond (Si), and Hexagonal Closed Pack (HCP-Mg) with 40 grains. For the prediction of each of the aforementioned classes, we generate 10 polycrystalline training samples (see supplementary Fig. S6 (a-d)) with atomsk[66] package. The overall characterization is carried out with a two-step approach. First, we label the atoms locally based on their crystalline motifs (e.g, FCC, BCC, *etc*.) and then, we apply an unsupervised learning DBSCAN[67,68] clustering to identify the size of the grains in the polycrystal samples. The grain size distribution and the number of particles belonging to crystalline motifs as predicted by CEGANN and CNA have been compared in Fig. 5. It is to be noted that the ordinary CNA cannot distinguish diamond structure. Hence, we use a modified CNA[69] for the creation of the labels of the Si (diamond structures). The number of nearest neighbors used for the construction of the graphs for each of the classifications is reported in Table. 1. This conforms to the number of neighbors that traditional CNA uses[31] for the prediction tasks.

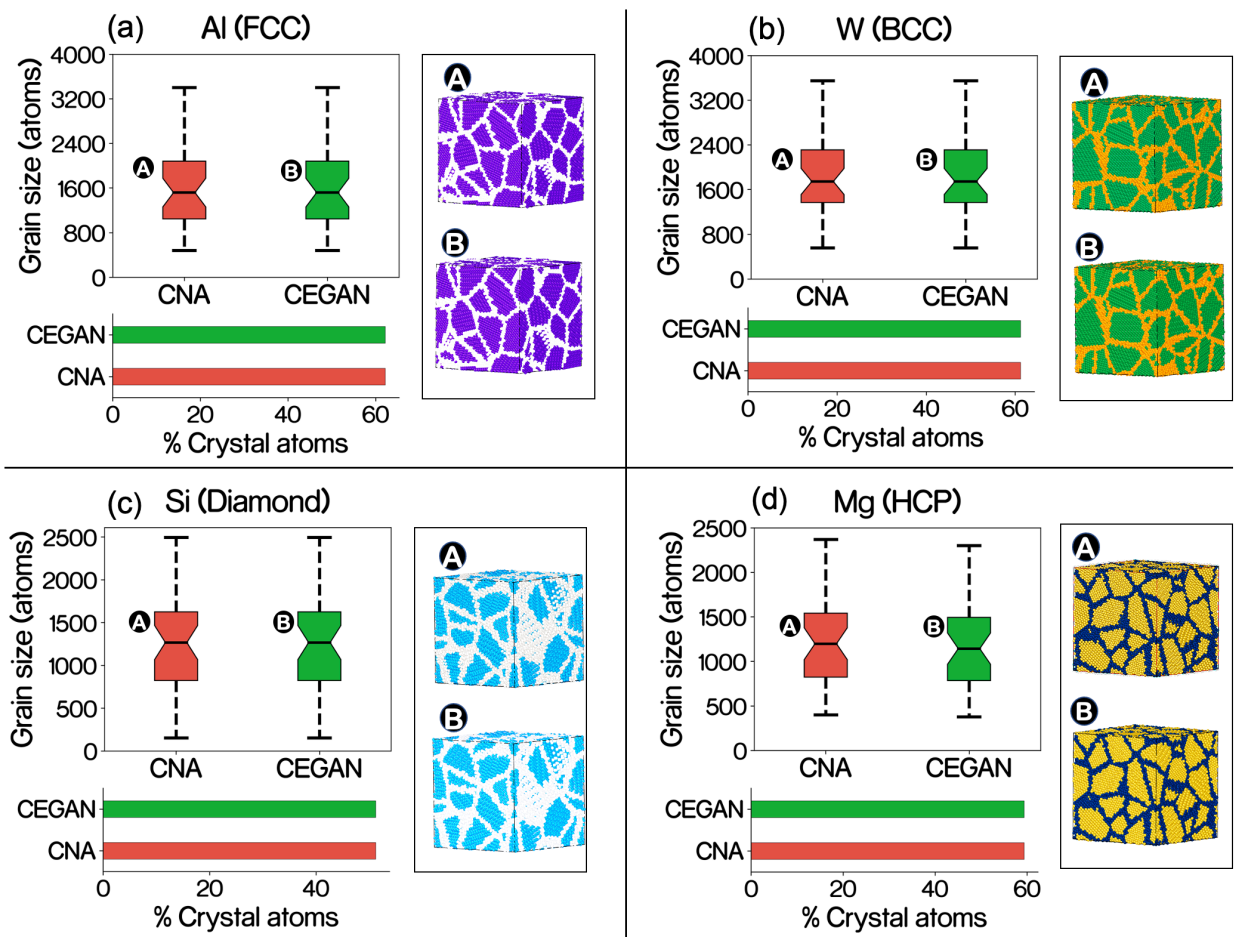

*Fig. 5.* *Grain size distribution of polycrystals of (a) Aluminum (FCC), (b) Tungsten (BCC), (c) Silicon (Diamond), (d) Magnesium (FCC) computed using CEGANN+DBSCAN clustering and CNA (Common neighbor analysis) + DBSCAN clustering.*

The predictions of CEGANN (Fig. 5 (a-d)) are almost identical to those of CNA, both in terms of the grain size distribution and the number of particles belonging to crystalline motifs of the gains. This clearly demonstrates the ability of the CEGANN in learning the different local motifs and distinguishing them from disorder atoms. The predictions of CEGANN on the local level classification tasks are largely dependent on the selection of the number of convolutional layers in the model as well as the number of neighbors used for the local neighborhood of the edge graphs. Adding more convolution layers will cause the compression of too much information at a single node. This may result in a loss of resolution, which in turn would deteriorate the CEGANN performance. As we increase the number of convolutional layers for fixed 12 neighbors of graph construction (Fig. 6 (a)), the performance severely declines at 4 edge-convolutional (+3 angle convolutions) layers. However, it seems that with an increase in the number of neighbors, CEGANN tends to slightly underpredict grain sizes (Fig. 6 (b)). The amount of information being compressed in each node of a graph using subsequent convolutions follows the equation:

$$N_{\text{infomation}} = NN^{\text{CONV}} \quad (1)$$

where, $N_{\text{infomation}}$ is the information from surrounding neighbors in terms number of atoms, "$NN$" is the number of nearest neighbors of an atom in the graph, and "CONV" is the number of convolutions being used. The mean grain size of the Mg (HCP) system is ~1200 with a maximum value of ~2500. In Fig. 6(c), beyond the operation point 12, 3 ($NN$, CONV) the amount of information being compressed is ~8000, which is much larger than the maximum grain size. Hence, there is a severe mix-up between the information of grain boundary and grains. Thus, the model tends to perform poorly at 4 edge-convolutional (+3 angle convolutions) (Fig. 6 (a)). An increase in $NN$ will cause this deterioration very slowly and will result in an underprediction of grain sizes (Fig. 6 (b)). It is also worth mentioning that, unlike CNA, CEGANN is very flexible in learning environments with local noises, such as thermal noise, which is essential for practical applications.

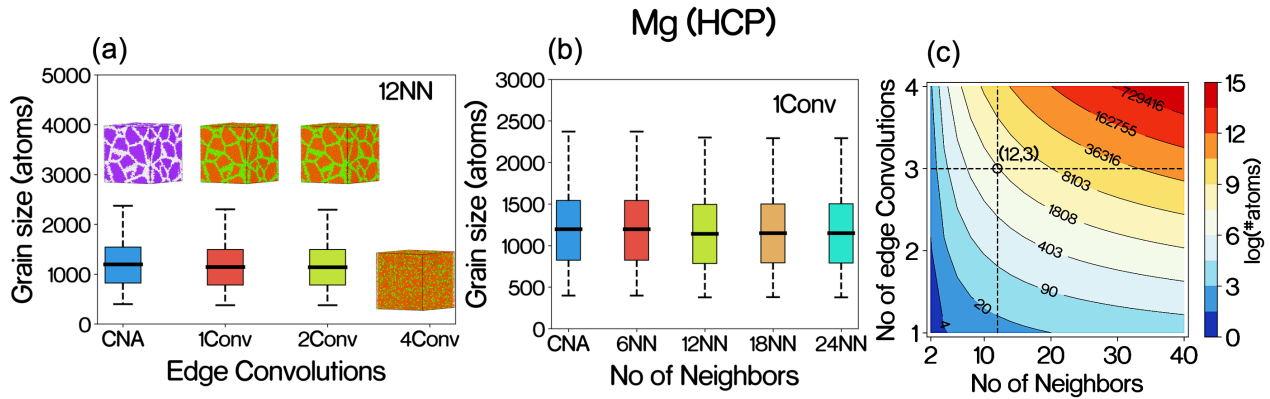

*Fig. 6.* (a) Effect number of edge-convolutional layer on the prediction of grain size distribution (b) Effect of number nearest neighbors used for graph construction on grain boundary prediction. (c) The amount of information (in terms of the number of atoms) being compressed in a node of the graph for different edge convolutions (& n-1 angle convolutions) and the number of neighbors used for the graph construction.

**Dynamical classification of structures with thermal noise:**

Zeolites are ordered microporous silicates or [70,71]aluminosilicate materials widely used as solid catalysts in the chemical industry.[70,71] Knowledge about the mechanistic pathways of the formation of zeolites is still limited, and the key to realizing new zeolites for catalysis and separations. The stochastic nature of nucleation processes and the small, nanoscopic size of critical nuclei within the heterogeneous reaction mixture, make the detection of the birth of a new phase challenging in experimental hydrothermal synthesis. Molecular simulations have the right spatial resolution. However, in the synthesis mixture, the zeolite crystallites and the surrounding amorphous matrix have very similar local and medium-range orders[72]. Fig. 7(a-b) shows that, indeed, the zeolite and the network former silica in the amorphous phase has very similar radial and $Q^n$ (number of silica neighbors) distributions. Moreover, unlike the case of simple crystals, such as ice, where the unit cell consists of 1-2 atoms, the unit cell of zeolites typically has ~100 silica nodes. Even though each silicon has a coordination number of 4, the environment of each silicon node is diverse in the zeolite. This makes the identification of the nascent zeolite inside an amorphous matrix a very challenging endeavor.

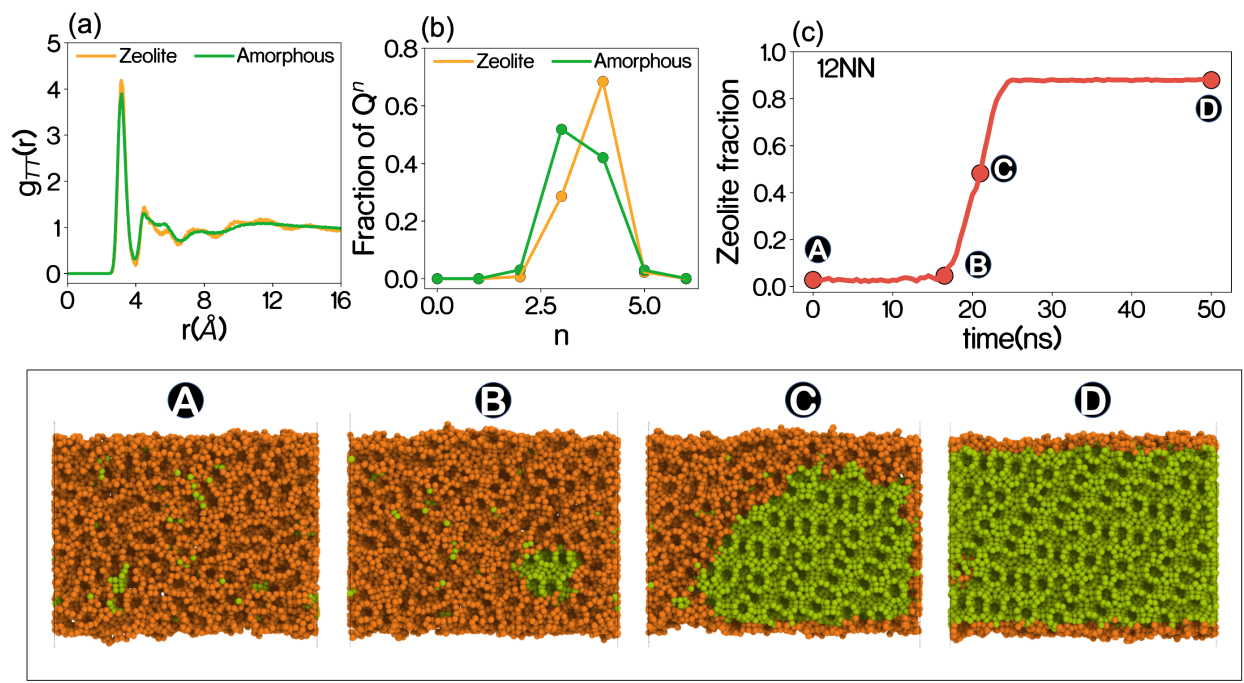

*Fig. 7. Performance of CEGANN for dynamical classification during the process of nucleation and growth of a zeolite crystal from an amorphous parent phase. (a-b) The radial distribution function between silica nodes ($g_{T-T}$) and the number of silica neighbors ($Q^n$) is very similar between the amorphous and zeolite phases. (c) CEGANN predicts the fraction of silica sites that are part of the zeolite, as it nucleates and grows from the synthesis mixture. The snapshots of the simulation box corresponding to points A-D are shown in the lower panel. Silica nodes of the amorphous phase are shown in orange, whereas the crystalline*

*silica detected by CEGANN is shown with green. Fort clarity, the organic cations and water molecules are not shown.*

Traditional approaches, such as the bond-orientational order parameter $q_6$, could be used to detect the nucleation process of zeolites. However, the requirement of the large cutoff distance makes it inefficient to detect very small nuclei[73–75]. Moreover, the bond-order parameter approach is specific to a particular zeolite polymorph. Identification of crystal based on mobility criteria is not zeolite specific, but it assumes that there is a considerable mobility difference between the new crystal phase from the mother phase[76]. This approach does not work if the new phase crystallizes from a glassy state, as is the case in zeolite synthesis[72]. These necessitate the development of a classification technique that distinguishes the zeolite nucleus from the amorphous phase during the formation of zeolites.

We use the CEGANN framework to probe the evolution of the zeolite nucleus and growth in the simulation mentioned above. To train our network we use a total of 400 structures consisting of 50% pure crystalline zeolites at different temperatures, noisy zeolite crystals (added Gaussian noise to the atomic positions) as well as 50% amorphous structures at different temperatures (see supplementary section S.4). We use 12 NN (nearest neighbors) (see Table. 1) for the graph construction and although the effects of 4 and 8 nearest neighbors on the construction of the graph are also explored (see supplementary Fig. S7 (c)). Fig. 7 (c) shows the zeolite fraction in the simulation trajectory as a function of time for the case of 12NN. A sharp change in the fraction of zeolite starting at time 16.5 ns suggests the formation of stable nuclei of zeolite Z1 that grow into a full slab at time > 25 ns. The same is evident from the snapshots presented at different instances during the crystallization (panel A-D in Fig. 7). This case study clearly illustrates that the proposed CEGANN workflow is not only capable of performing accurate classification in static local environments but also equally effective in heterogeneous simulation environments with considerable thermal noise. This is remarkable because our CEGANN workflow can identify a crystal nucleus smaller than its unit-cell size.

**Multilabel characterization of mesophases in binary mixtures:**

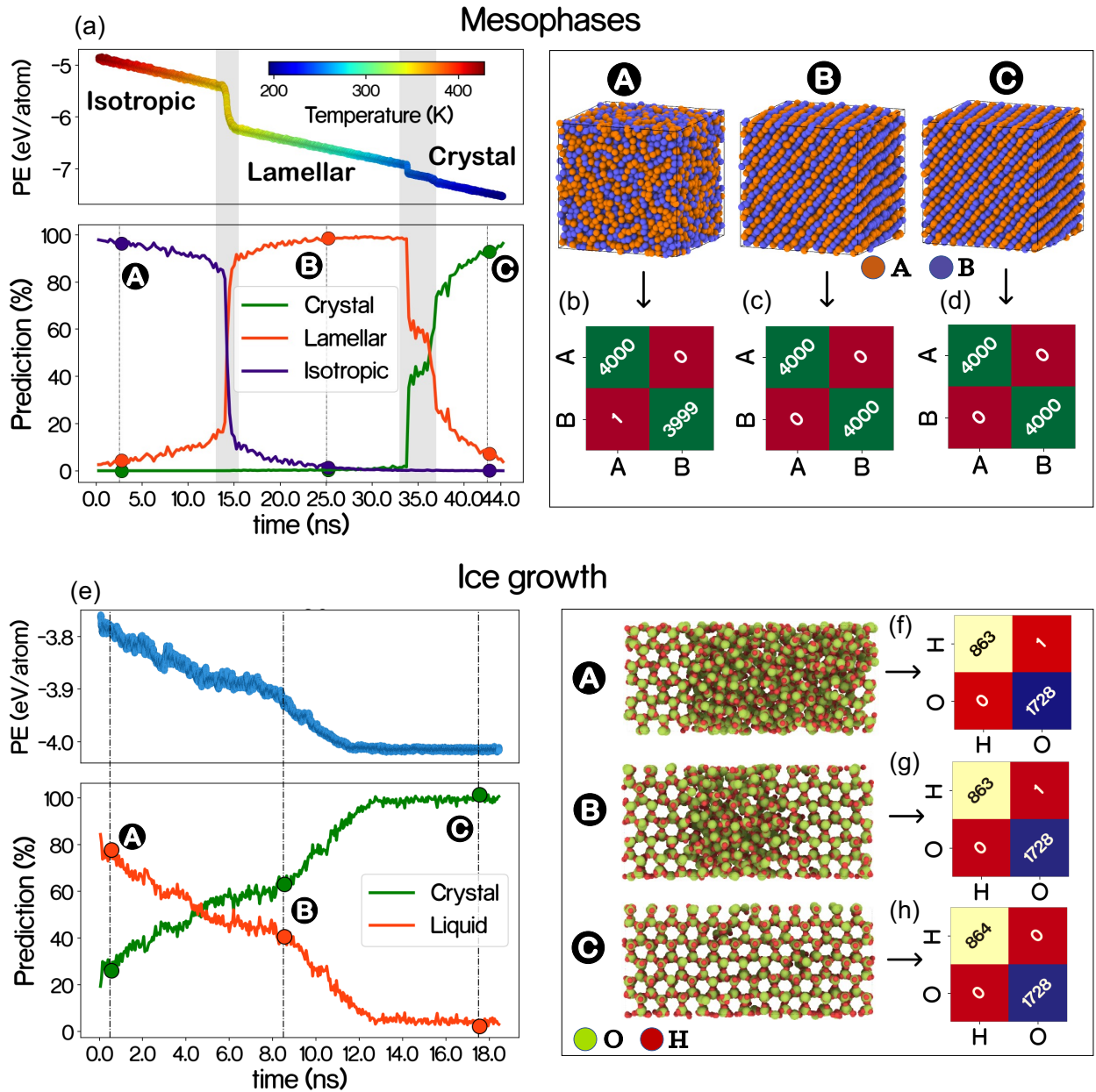

*Fig. 8. Multilabel characterization using CEGANN (a) Characterization of isotropic liquid, lamellar mesophase, and layered crystal in a binary synthesis mixture. CEGANN identifies each phase along thermal trajectories of phase transformation. Respective potential energy changes with time and corresponding predictions of existing phases by CEGANN. (b-d) are the confusion matrix of the overall chemical species predicted by CEGANN at different time steps (A, B, C) for mesophase characterization (e) Predicting the growth of ice from liquid at 235 K, along with the atoms ("H" & "O") present in the system (Multilabel). (f-h) are the confusion matrix of the atoms predicted at different time steps (A, B, C) as the system is crystalizing.*

Mesophases have order intermediate between that of amorphous and crystalline phases, and are traditionally observed in block copolymers and solutions surfactants, but can also occur in other systems with frustrated attraction.[77,78]. Mesophases occur on multiple morphologies such as lamellar, gyroid, and hexagonal[77,79] .The intermediate nature of the ordering in mesophases makes them challenging to identify in simulations. We use CEGANN workflow to characterize the formation of mesophases, and subsequent crystallization during the cooling of a binary isotropic mixture of representative species A and B[77]. We also characterize the order of the species in the system as the phase transition is taking place. The data set consisted of 22 lamellar, 22 crystalline and, 22 isotropic mixture structure (see supplementary information section S.5 and Fig. S8). The transitions are validated with the potential energy changes in the system (side panel of Fig 8(a)). Fig. 8 (a) demonstrates that CEGANN successfully characterizes the amorphous, lamellar and crystalline phases individually, and also accurately detects the transition between phases along a cooling simulation.

**Multilabel classification of interface evolution during ice growth:**

The crystallization of water is ubiquitous in natural environments. Development in the last two decades, have enable simulations of ice nucleation and growth with molecular resolution[81–84]. Here, we implement CEGANN for the characterization of early stages of growth of ice I from liquid water, a polyatomic molecule. The simulation was carried out using the TIP4P/2005[85] water model (see supplementary information section S.5 and Fig. S8). Using our multilabel classification approach, we classify whether a particle belongs to either crystalline or liquid phase, and also identify the local order of each water molecules. In Fig. 8 (e), we show that CEGANN precisely characterizes the crystallization of water (reflected in the decrease of the potential energy of the system).

The above two examples –phase transitions in the binary mixture and crystallization of water- display CEGANN's ability to precisely characterize complex environments with multiple components or polyatomic species, in the presence of thermal noise.

**Discussion**

Characterization of materials at different scales and domains of application is a must for any data-driven material science application. In this work, we develop the graph attention based CEGANN workflow, which is transferrable across scales and adaptable to variabilities in the material environment, while also providing an accurate characterization. We demonstrate the efficacy of our workflow by taking on challenging and yet relevant classification problems in material science. Unlike other graph based architecture (CGCNN) or mathematical formulation-based descriptors (SOAP, order parameters), CEGANN is not only able to classify disordered (liquid and amorphous phases) at a global level but equally accurate in classifying local motifs in stacking disordered structures, displaying transferability in the application domain. It is equally effective in performing global level classification tasks such as space group classification and characterization of structures based on their dimensionality. These play crucial roles in a plethora of material science problems. CEGANN performs the classifications with near-perfect accuracy while its counterparts (SOAP, CGCNN, etc.) fail to do so.

We further demonstrate that CEGANN successfully addresses challenging problems. It characterizes the grain boundaries in polycrystalline materials and their grain distribution in terms of size, with identical accuracy in terms of grain size distribution to the traditional CNA method. It can identify the formation of complex crystals with large unit cells, identifying of onset of nucleation and growth of a zeolite from a synthesis solution with strong thermal fluctuations, even when the size of the nucleus is much smaller than the unit cell of the zeolite, and also captures accurately the growth process. We demonstrate the CEGANN can identify crystalline and amorphous phases in polyatomic systems with thermal noise, as well as distinguish liquid, mesophase and crystalline order in binary mixtures. These applications showcase the applicability of CEGANN to problems involving variability in the environments. Overall, our approach is agnostic to the problem and allows the classification of features at different scales with equal efficacy.

**Methods:**

**Angle convolution:**

In the angle convolutional layer uses bond angle ($\theta_{ijk}$) cosines expanded on a gaussian basis as the initial input. The idea is that each angle learns and collects the messages from its adjacent edges through the convolutions. We use a simple graph attention-based architecture and convolutional operation is performed according to

$$\boldsymbol{\theta_{ijk}^{l+1} = softplus\left(\theta_{ijk}^{l} + \alpha_{ijkl} * \left(W_{ijkl}^{f}(\theta_{ijk}^{l} \oplus e_{ij}^{l} \oplus e_{jk}^{l}) + b_{ijkl}^{f}\right)\right)}, \quad (1)$$

where $e_{ij}^l, e_{jk}^l$ are edge features from previous edge convolution layers an $\alpha_{ijkl}$ is the attention coefficient calculated using[86]

$$\boldsymbol{\alpha_{ijkl} = softmax\left(\left(W_{ijkl}^{att}(\theta_{ijk}^{l} \oplus e_{ij}^{l} \oplus e_{jk}^{l}) + b_{ijkl}^{att}\right)\right)}, \quad (2)$$

where $W_{ijkl}^{f}, W_{ijkl}^{att}$ and $b_{ijkl}^{f}, b_{ijkl}^{att}$ are feature and attention weights and biases, respectively. We use softmax activation as a normalizer for calculating the attention coefficient and the final output of the angle convolution is passed through a softplus activation to obtain the final representation. Batch normalization is applied after the aggregation operation.

**Edge convolution:**

We follow a similar attention type mechanism for the edge convolutional layer. The convolutional function is represented as

$$e_{ij}^{l+1} = softplus\left(e_{ij}^l + \sum_{k \in N} softplus\left(\alpha_{ijk} * \left(W_{ijk}^f(\theta_{ijk}^l \oplus e_{ij}^l \oplus e_{jk}^l) + b_{ijk}^f\right)\right)\right), \quad (3)$$

where $W_{ijk}^f$ and $b_{ijk}^f$ are the weights and biases for the feature matrix and $\theta_{ijk}^l$ is the angle features from the previous angle convolutional stage. $\alpha_{ijk}$, the attention coefficient computed using an equation analogous to Eq.2, with different weights and biases. We apply a nonlinear $softplus$ activation function before and after the aggregation over the neighborhood; the additional nonlinearity helps the features to adapt to the target task. There is also provision of adding explicit one hot coded atomic feature $x_i$ based on the characterizing task being performed. The incorporation of the chemical information is done before each edge convolution. For "$l+1^{th}$" edge convolution layer, with "$e_{ij}^l$" as input form the "$l^{th}$" layer, the atomic features of atom "$i$" and "$j$" ($x_i, x_j$) are included as a concatenation of the features (Fig. 2 (c)).

$$e_{ij}^l = e_{ij}^l \oplus x_i \oplus x_j$$

**Feature aggregation and concatenation:**

The aggregation block (Fig. 2(c)) consists of 3 stages. First, the edge and angle features are aggregated as

$$e_i^{l+1} = \sum_{j \in N} softplus(e_{ij}^l) \quad (4)$$
$$\theta_i^{l+1} = \sum_{j \in N} softplus\left(\sum_{k \in N} softplus(\theta_{ijk}^l)\right) \quad (5)$$

The final feature representation is given as concatenation $Z_i = e_i^{l+1} \oplus \theta_i^{l+1}$. To provide more resolution to the aggregated feature, we take a linear transformation before the aggregation stage. The pooling of the features follows the concatenation operation. It should be noted that the pooling (average-pooling) on the features is applied only if a global level classification task is being performed. For local classification tasks, no pooling is applied to the features. Batch normalization is applied after the aggregation operation. We also apply dropouts' (0.5 rate) before subsequent transformation after the convolutional layer. This helps in reducing overfitting. We use cross-entropy loss as the loss metric[87].

**Training the model:**

The network is trained on 1 GPU-accelerated to compute node on the NERSC computing cluster with 20-core Intel Xeon Gold 6148 ('Skylake') @ 2.40 GHz and 1 NVIDIA Tesla V100 ('Volta') GPU. The feature vector for the Angle convolution and edge convolutions are kept being 80. The hidden feature for the dense layer following the edge and angle convolution layers is 256. Upon aggregation, the overall dimension of the feature vector is 512.

**Code & Data Availability:**

The dataset used for space group classification is available from https://www.nomad-coe.eu/[88]. The Carbon bulk structures used in this work are available in https://www.sacada.info/. The CEGANN code along with the trained models and datasets are available via https://github.com/sbanik2/CEGANN.

**Acknowledgment**:

The authors acknowledge support by the U.S. Department of Energy through BES award DE-SC0021201. Use of the Center for Nanoscale Materials, an Office of Science user facility, was supported by the U. S. Department of Energy, Office of Science, Office of Basic Energy Sciences, under Contract No. DE-AC02-06CH11357. This research also used resources of the Argonne Leadership Computing Facility at Argonne National Laboratory, which is supported by the Office of Science of the U.S. Department of Energy under contract DE-AC02-06CH11357. This research used resources of the National Energy Research Scientific Computing Center; a DOE Office of Science User Facility supported by the Office of Science of the U.S. Department of Energy under Contract No. DE-AC02-05CH11231. We gratefully acknowledge the computing resources provided on Fusion and Blues; high performance computing clusters operated by the Laboratory Computing Resource Center (LCRC) at Argonne National Laboratory.

**Author contributions**

S.B. and S.K.R.S conceived the project. S.B. developed the CEGANN workflow for multiscale classification. H.C provided feedback on the workflow. S.B evaluated the performance of workflow on different classification tasks and analyzed the results. SB wrote the manuscript with guidance from S.K.R.S and V.M. D.D and V.M generated the dataset and contributed to the writing, and analysis of the dynamical classification, Mesophase and Ice growth characterization task presented in the manuscript. S.M, H.C, and D.D provided feedback on the manuscript. M.C. assisted with the computing resources. All authors participated in discussing the results and provided comments and suggestions on the various sections of the manuscript. S.K.R.S supervised and directed the overall project.

**Competing Interests**

The Authors declare no Competing Financial or Non-Financial Interests.

# Supplementary Information

# CEGANN: Crystal Edge Graph Attention Neural Network for multiscale classification of materials environment

Suvo Banik[1,2], Debdas Dhabal[3], Henry Chan[1], Sukriti Manna[1,2], Mathew Cherukara[4], Valeria Molinero[3], Subramanian KRS Sankaranarayanan[1,2]

[1] Center for Nanoscale Materials, Argonne National Laboratory, Lemont, Illinois 60439, United States.
[2] Department of Mechanical and Industrial Engineering, University of Illinois, Chicago, Illinois 60607, The United States.
[3] Department of Chemistry University of Utah Salt Lake City, UT 84112, United States.
[4] Advanced Photon Source, Argonne National Laboratory, Lemont, Illinois 60439, United States.

## S.1. Generation of amorphous and liquid silicon structures:

To generate the liquid and amorphous (glassy) structure, pure cubic diamond silicon is melted at 6000 K using an NPT ensemble with pressure ranges 0.1-4 GPa, which gives the liquid dataset. The atomistic simulations are carried out using a Stillinger-Weber[1] interatomic potential with a LAMMPS package. To create glassy amorphous structures, we quench the liquid structures at temperatures within 50-350 K (below the glass transition of silicon) using an NVT ensemble. The typical amorphous structure has a sharp radial distribution function (rdf) peak at ~2.4 Å and a secondary peak at ~4 Å as compared to the liquid, where the distribution is flat beyond the first peak (Fig. S1 (a)). Despite the difference in the rdfs, they both take similar volume per atom (Fig. S1 (b)).

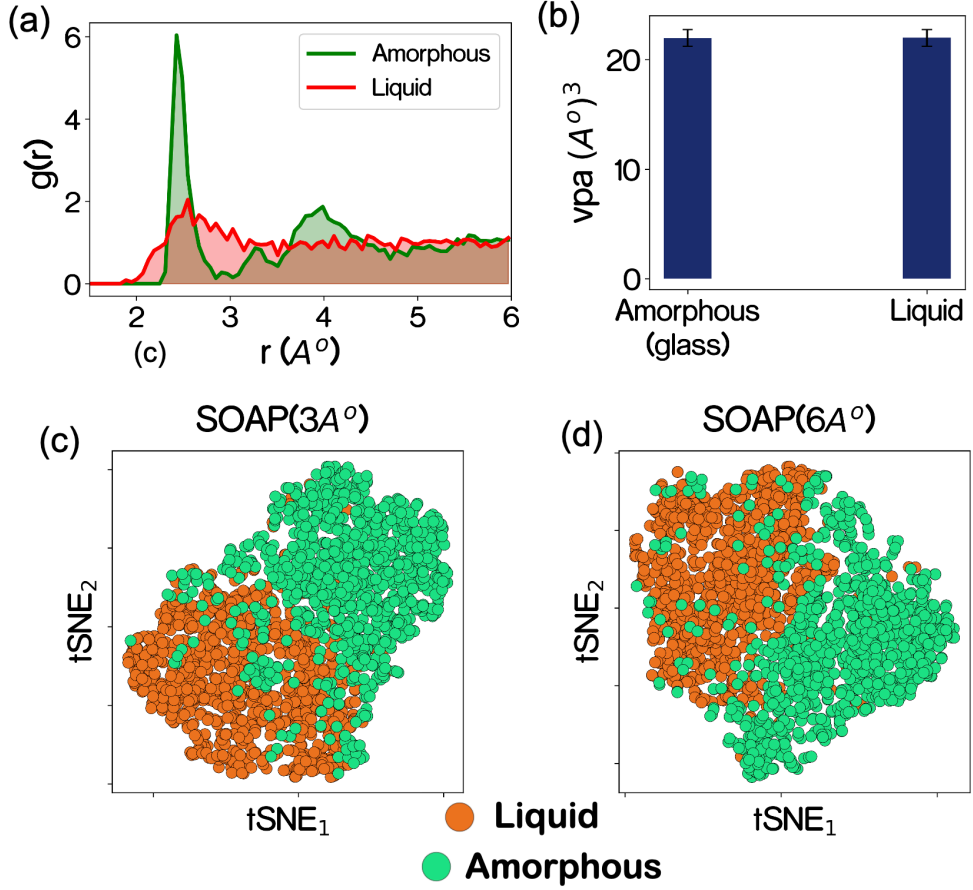

*Fig.S1.* *Classification of liquid and amorphous Si structures modeled using Stillinger Weber (SW) potential. (a) Comparison of radial distribution functions between liquid and amorphous structures. (b) Volume per atom (vpa) of liquid and amorphous silicon. PCA plot of SOAP representation of the liquid and amorphous phases for a cutoff of (c) 3 Å and (d) 6 Å on the overall dataset.*

## S.2. Additional chemical features inclusion for muticompoenent system:

The incorporation of the chemical information is done before each edge convolution. For "$l+1$" th edge convolution layer, with "$e_{ij}^l$" as input form the *"l"* th layer, atomic features of atom *"i"* and *"j"* ($x_i, x_j$) are included as a concatenation of the features. (Fig.S2 (a)).

$$e_{ij}^l = e_{ij}^l \oplus x_i \oplus x_j$$

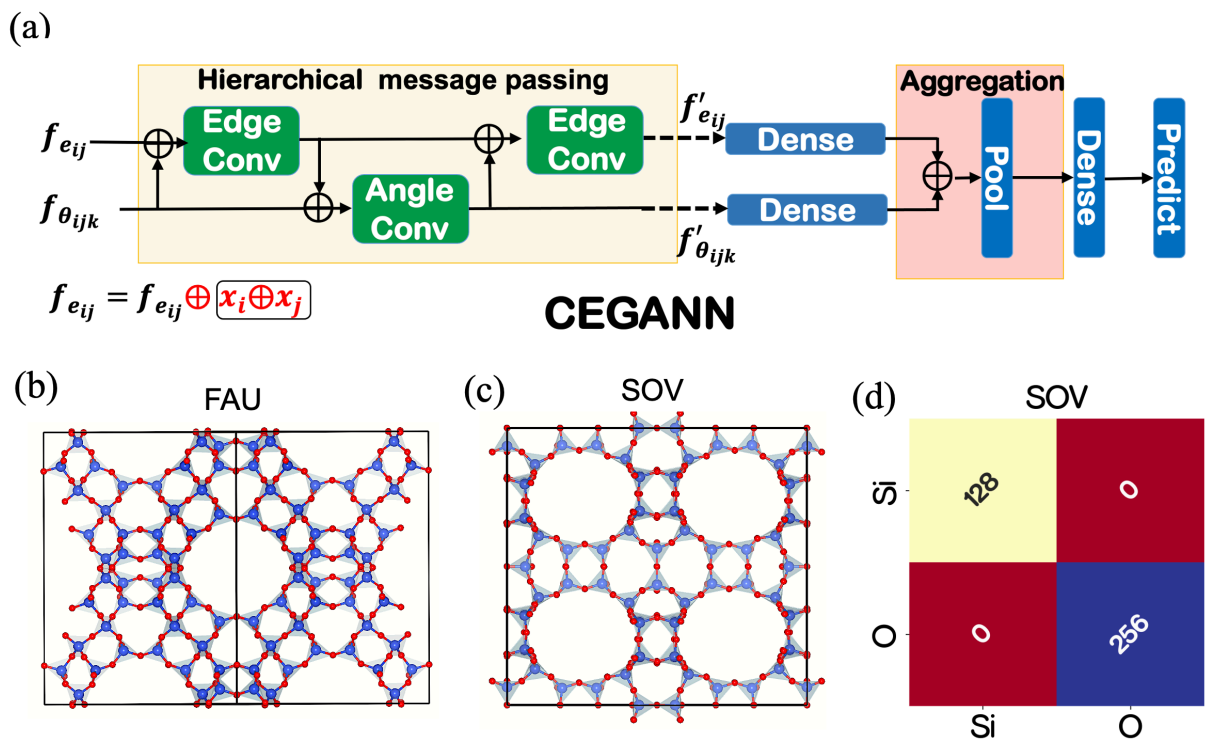

***Fig. S2***. *(a) CEGANN workflow incorporating the chemical information of the species at the aggregation stage. (b) FAU framework (c) SOV framework (c) Confusion matrix of the prediction of CEGANN on SOV framework on characterizing "Si" and "O" chemical species.*

A simpler way to generate the atomic features is through the one-hot encoding of the species present in the system. For example, a ternary system with "C-H-O" species will have one hot encoding [1,0,0] for the "C" atom, [0,1,0] for the "H" atom, and [0,0,1] for the "O" atom. It is also worth mentioning that, apart from including explicit chemical information, CEGANN also learns the distinction between the atomic species from the interatomic distances of surrounding atoms, which constitutes the local environment of it in an edge graph representation. To demonstrate this, we conducted a test by characterizing the "Si" and "O" in two distinct frameworks of zeolite, "FAU" and "SOV" respectively without including any chemical information in the network beforehand. We train CEGANN on the "FAU" framework and predict the species "SOV" framework (Fig.S2 (b-c)). 12 nearest neighbor and 1 convolution were used for the construction of the graph and training respectively. From the confusion matrix (Fig. S2 (d)), it can be seen CEGANN identifies the species with perfect accuracy.

Hence the choice to include explicit chemical information in the network depends on the characterization task being performed. For systems with single species, it is not needed. For multinary systems, inclusion of chemical information might be necessary. Inclusion of these additional features, however, increases the network parameters and thus the cost of training. Since CEGANN has an inherent ability to learn these features without additional information, in the end it becomes a tradeoff between cost and accuracy – the selection depends on the user who can utilize the multilabel classification scheme in CEGANN depending on the application of interest.

### S.3 : Selection scheme of hyperparameters for the model:

A simple grid search approach is followed to obtain the optimal values of the two model hyperparameters "No of Edge Convolutions" (Which also defines the "Number of angle convolutions") and the "Number of nearest neighbors used for the construction of the graph". Fig. S3 (a-b) shows the grid search results for the local characterization (stacking-disordered ice) and global characterization task (structural dimensionalities 0D, 2D, 3D etc.) Fig. S3 (c) shows the total number of trainable parameters of the model varying with the above-mentioned parameters. It is pretty evident that the number of trainable parameters is very sensitive towards the number of convolutions used (which directly relates to the computational cost associated) while the selections of the "Number of nearest neighbors" is very much problem specific. For the local classification task, the accuracy is good only for parameters with 16, 12, 20 nearest neighbors and 1 edge convolution (Fig. S3 (a)). This also relates to the physical intuition that information of the $2^{nd}$ nearest neighbors is needed for the hexagonal and cubic diamond structure characterization[2]. For the global classification task, the accuracy is not much dependent on the "Number of nearest neighbor used" for the graph construction, instead, it degrades at a higher "Number of edge convolutions" primarily due to computational complexity (higher number of trainable parameters (Fig. S3 (c)) associated with the training. Hence the choice of a good metric lies between a moderate to low associated training cost with good accuracy of the model. For the local characterization, a value of 1 edge convolution and for the 16 nearest neighbors and the global characterization 2 edge convolution and the 12 nearest neighbors have been selected.

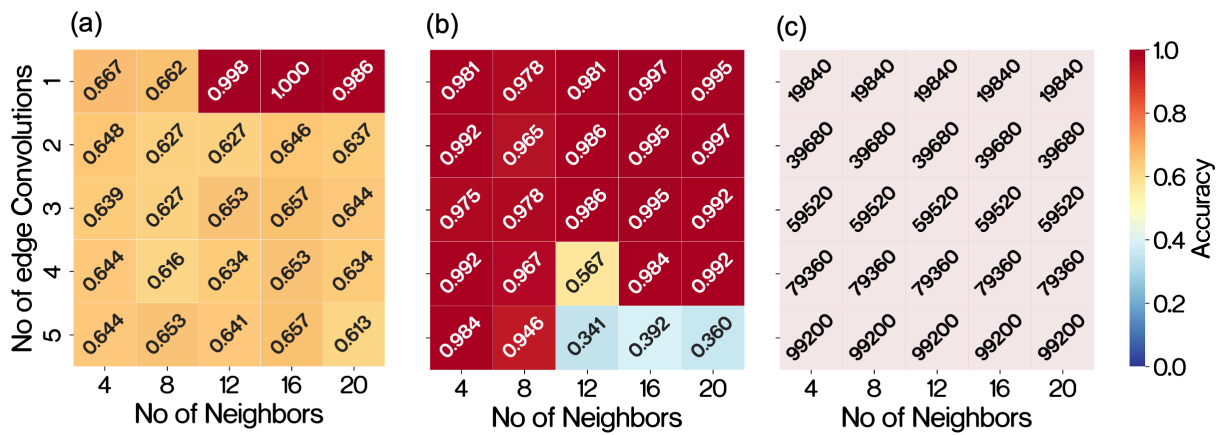

*Fig. S3.* *(a) Prediction accuracy for the characterization of stacking disordered ice for different sets of model hyperparameters used. (b) Prediction accuracy for the characterization dimensionality characterization task. (c) Variation of the total number of trainable model parameters for different sets model hyperparameters used.*

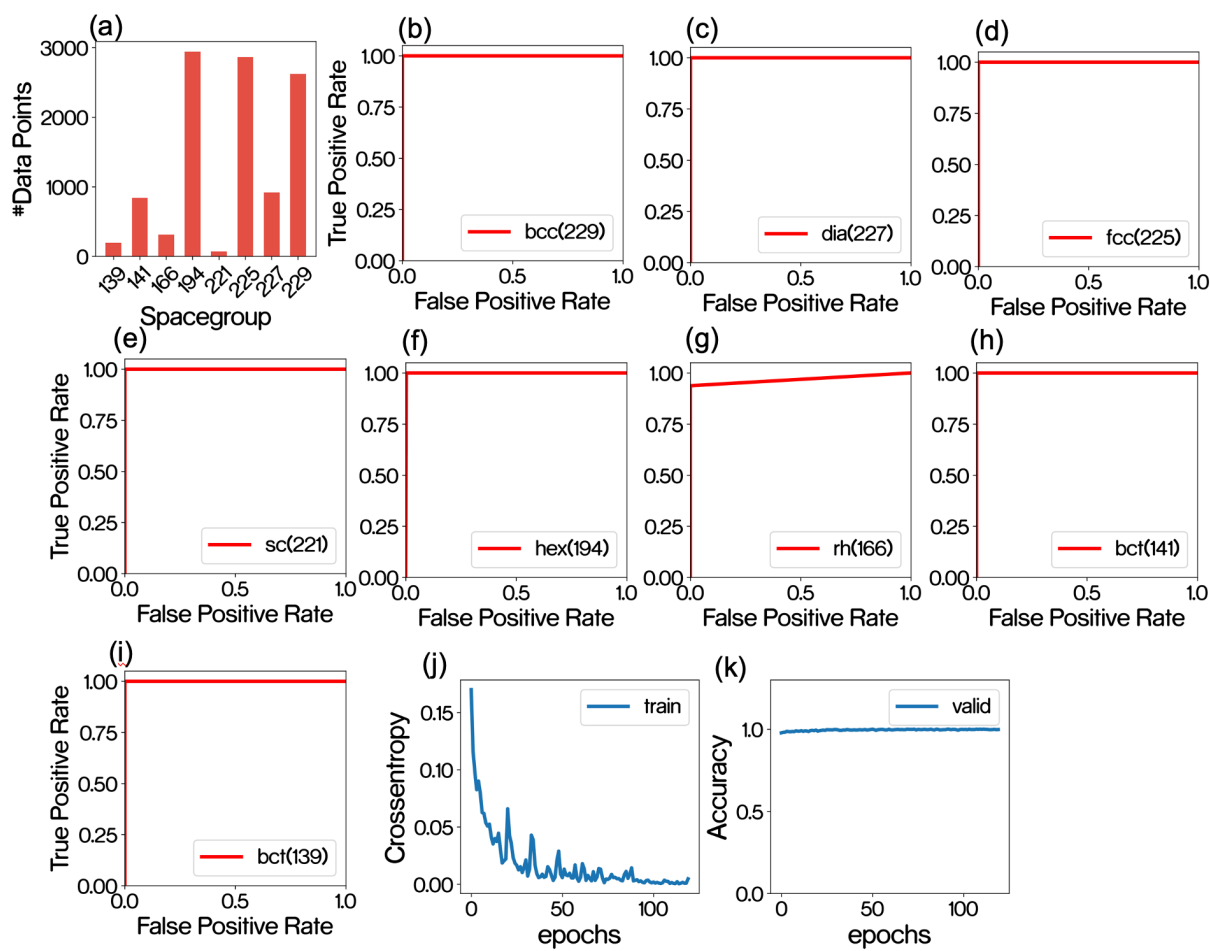

*Fig. S4. (a) The data points in each class in the dataset used for space group classification. (b-i) The ROC-AUC curves as obtained on the validation datasets results from the CEGANN workflow. (j-k) Typical training and validation accuracy during the training of the CEGANN for the classification task performed.*

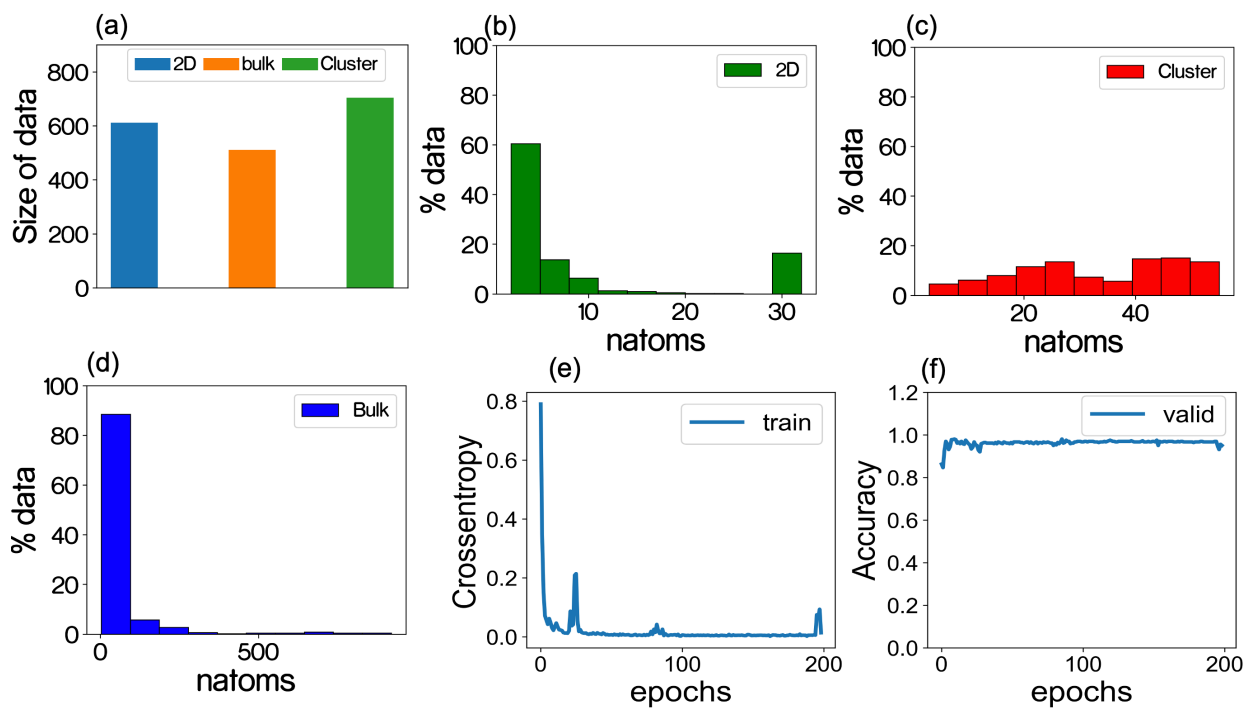

*Fig. S5.* Dataset used by CEGANN for dimensionality classification. (a) The data points in each class in the dataset used. (b) Size distribution of 2D polymorphs (c) Size distribution of clusters (d) Size distribution of bulk structures used in this study. (e) Cross entropy loss during training on the training set. (f) Accuracy score on the validation dataset during training.

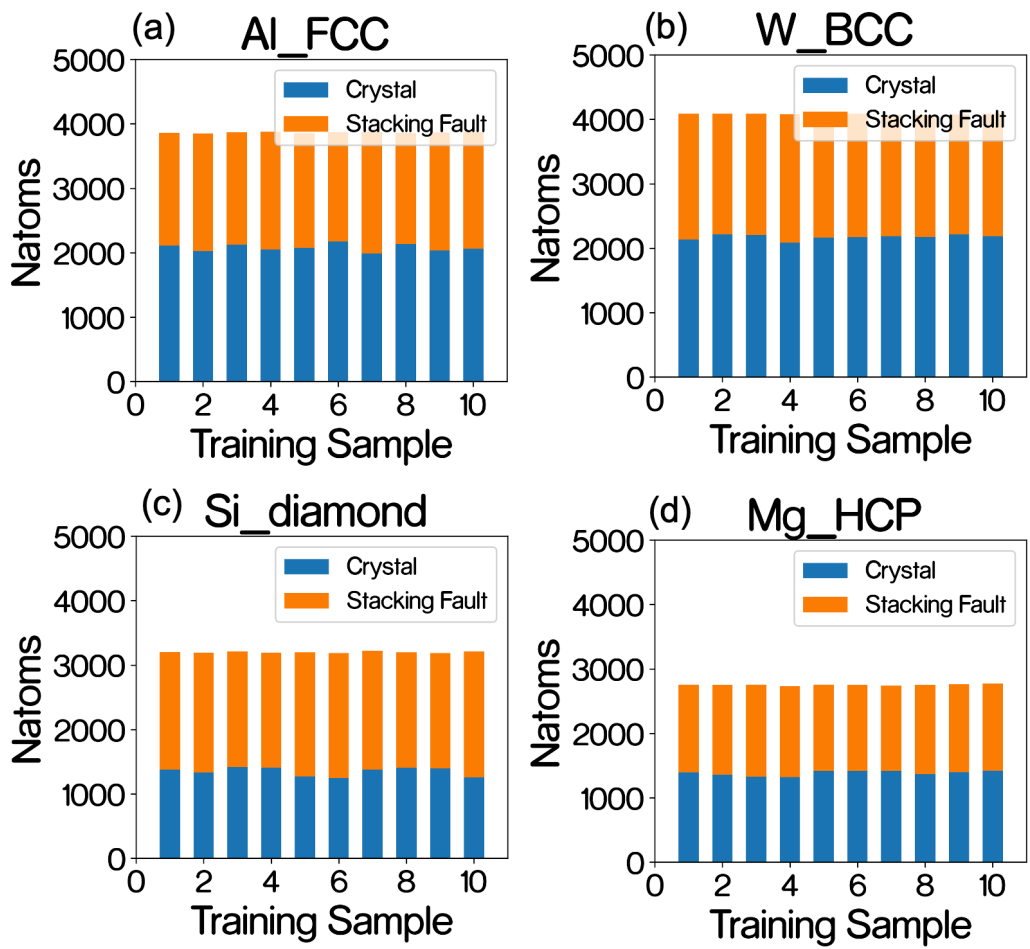

***Fig.S6****. (a-d) Crystalline and stacking fault particles in training data for the Al(FCC), W(BCC), Si (diamond), and Mg (HCP) system.*

**S.4. Generation of data for probing nucleation and growth in zeolites:**

The prediction data used for the identification of Z1 zeolite in this study has the structure of the FIR-30 metal-organic framework[3]. We select a molecular dynamics simulation trajectory at 670 K that depicts the nucleation and growth process simulated using a coarse-grained model[4]. The overall training data contains pure zeolite Z1 and the amorphous silica phase. We use both, energy minimized zeolite structure and equilibrated framework at 670 -720K for the training purpose. Similar is the case for the amorphous phase too, except that is equilibrated at. We observe a significant fluctuation in the high-temperature zeolite structure even though the microscopic structural environments are the same between the low and high-temperature states (Fig. S7 a-b). To account for the noises in the nucleated crystal, we consider pure zeolites with Gaussian noise added to their atomic positions. Overall, the training data contains a diverse set of crystal and amorphous phases that helps the network to learn beyond thermal fluctuations and identify critical features of nucleation and growth.

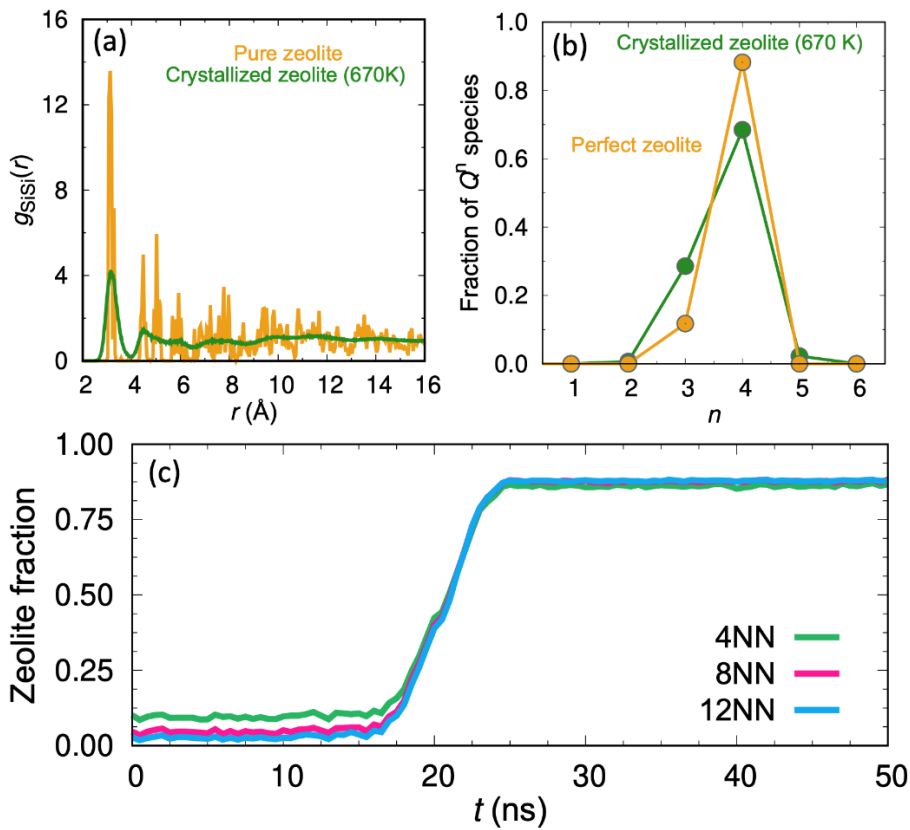

***Fig. S7.*** *(a) The rdfs of the pure energy minimized zeolite crystal and a zeolite at a temperature of 670 K used for the prediction. (b) The coordination number distribution, $Q^n$ of the same. (c) The fraction of nucleating Z1 particles and its growth predicted by CEGANN for different number of neighbors.*

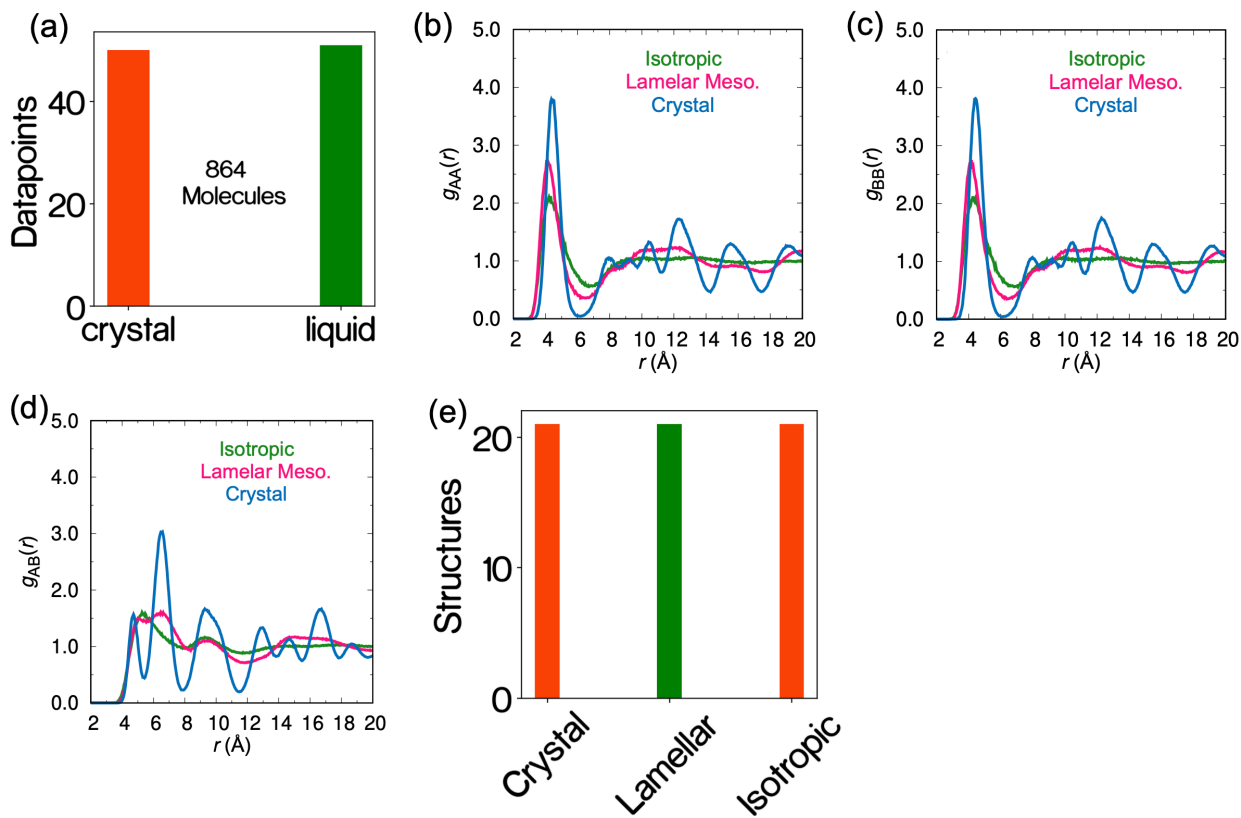

*Fig. S8.* (a) Distribution of the dataset used for the characterization of Interfacial growth of ICE (b-d) Partial rdfs for the lamellar, isotropic and crystalline phases for characterization of mesophases in binary mixtures. (e) The distribution of the dataset used for the same.

**S.5. Generation of data for Mesophase characterization and Multilabel classification of interface evolution during ice growth:**

**Ice Growth Simulations:**

Ice growth simulations are performed using TIP4P/2005 water model[5] in the LAMMPS package[6]. Standard interaction parameters for TIP4P/2005[5] model are used. The equation of motions is integrated using the velocity-Varlet algorithm with a 1 fs timestep. The simulation cell is periodic in three Cartesian directions. Both the fluctuations in temperature and pressure in NPT ensemble are controlled using Nose-Hoover thermostat and barostat with a time constant of 1 ps and 10 ps respectively. The bonds and angles in water molecules are constrained using the SHAKE algorithm[7]. The cutoff distances for short-range electrostatic and Lennard-Jones interactions are 10 and 12 Å, respectively. The long-range electrostatic interactions are calculated using particle-particle particle-mesh (PPPM) solver.

We first generate a rectangular ice slab with dimension ~23.5 Å × 22 Å × 54.2 Å containing 864 water molecules and minimize the energy at 1 bar with anisotropic coupling in the three dimensions using the box/relax command in LAMMPS. The minimized ice slab is then equilibrated at 235 K temperature. We then melt 80% of the ice slab keeping a 1.2 nm layer frozen to prepare a solid-liquid interface. We then perform a 20 ns long NPT simulation at 235 K to see the growth of the ice. The simulation trajectories are saved every 50 ps for analysis.

We further generate an ice slab with 128 water molecules. The system is energy minimized and equilibrated at 235 K following the procedure mentioned above. We then perform a production simulation of 1 ns saving 50 trajectories. The same ice slab is melted at a higher temperature to get a water box. This water box is then equilibrated at 235 K for 2 ns. The equilibrated water box is then used for a production run of 1 ns where 50 simulation configurations are saved. These smaller simulation configurations (ice and water) are used for training the ML-algorithm.

**Simulations of Phase Transformation in Binary Mixture:**

The simulations are performed using the LAMMPs package. The equimolar isotropic mixture of A and B type particles considered in this study interacts via the two-body part of the Stillinger-Weber potential. The energy and length-scale parameters for the interactions between particles used in this study is same as the one presented in Figure 4 of Kumar *et al*[8]. A random configuration with 8000 particles ($x_A = x_B = 0.5$) is generated in a cubic simulation box using Packmol software package. The simulation box is first energy minimized and then equilibrated under *NpT* conditions at 420 K and 0 bar for 5 ns. We take the final trajectory of 420 K simulation and cool the system isobarically (0 bar) till 200 K with a ramping rate of 5 K ns$^{-1}$. The details of integration algorithm, timestep and barostat and thermostat time constants are same as that of Kumar et al[8]. A total of 176 [8]frames are saved in the cooling simulation and are used to perform the classification.

To generate liquid configurations for the training purpose, starting from an isotropic mixture we perform 5 ns simulations at 400 K. The last 22 configurations are used for the training. The same is followed for lamellar and crystalline phase but at 300 K and 200 K respectively.